\documentclass[a4paper,11pt]{article}
\pdfoutput=1 

\usepackage{jcappub} 

\title{\boldmath $R^2$-gravity quark stars from perturbative QCD}

\author[a]{José C. Jiménez,}
\author[b,c]{Juan M. Z. Pretel,}
\author[b]{Eduardo S. Fraga,}
\author[b]{Sergio E. Jor\'as,}
\author[b,d]{and Ribamar R. R. Reis}

\affiliation[a]{
 Instituto de F\'\i sica, Universidade de S\~ao Paulo,\\Rua do Mat\~ao, 1371, Butant\~a, 05508-090, SP, Brazil}
\affiliation[b]{Instituto de F\'\i sica, Universidade Federal do Rio de Janeiro, \\ CEP 21941-972 Rio de Janeiro, RJ, Brazil}
\affiliation[c]{Centro Brasileiro de Pesquisas F{\'i}sicas, \\Rua Dr. Xavier Sigaud, 150, Urca, 22290-180, RJ, Brazil}
\affiliation[d]{Observat\'orio do Valongo, Universidade Federal do Rio de Janeiro, \\ CEP 20080-090 Rio de Janeiro, RJ, Brazil}

\emailAdd{josec.jimenez91@gmail.com}
\emailAdd{juan04manuel91@gmail.com}
\emailAdd{fraga@if.ufrj.br}
\emailAdd{joras@if.ufrj.br}
\emailAdd{ribamar@if.ufrj.br}

\abstract{We investigate the structure of quark stars in the framework of $f(R)= R+ \alpha R^2$ gravity using an equation of state for cold quark matter obtained from perturbative QCD, parametrized only by the renormalization scale. We show that a considerably large range of the free parameter $\alpha$, within and even beyond the constraints previously reported in the literature, yield non-negligible modifications in the mass and radius of stars with large central mass densities. Besides, their stability against baryon evaporation is analyzed through the behavior of the associated total binding energies for which we show that these energies are slightly affected by the modified gravity term in the regime of high proper (baryon) masses.}

\begin{document}
\maketitle
\flushbottom

\section{Introduction}\label{Sec1}

Neutron stars provide unique opportunities for the investigation of matter under extreme conditions of density. For decades it has been theorized that cold quark matter might exist at the core of heavy neutron stars \cite{Buballa:2014jta} (see, e.g., \cite{Annala:2019puf} for a recent discussion) or even more surprisingly as strange quark stars \cite{Bodmer:1971we,Witten:1984rs,Alcock:1986hz,Haensel:1986qb}. Currently we have precise measurements of masses \cite{Demorest:2010bx, Antoniadis:2013pzd} and those of radii have been improving significantly recently \cite{Miller:2019cac,Riley:2019yda,Raaijmakers:2019qny}, both for maximal ($>2M_{\odot}$) and canonical ($\sim 1.4 M_{\odot}$) neutron stars. Nevertheless, such constraints are still unable to distinguish between predictions coming from strange stars and (hybrid) neutron stars.  The same is true in highly non-linear situations \cite{Zhu:2021xlu}, such as the one found in gravitational waves produced by binary neutron star mergers \cite{LIGOScientific:2016lio, LIGOScientific:2018dkp}. 

On the theory of gravity side, General Relativity (GR), with its Hilbert-Einstein action proportional to the Ricci scalar $R$, has proven to be successful in describing strong-field measurements found in black-hole collisions (e.g.~\cite{LIGOScientific:2016lio}). However, if applied to cosmological problems, such as the early-time inflation and late-time accelerated expansion of the Universe, GR is apparently insufficient (apart from adding a cosmological constant) \cite{Starobinsky1980, Capozziello2002, Carroll2004, Nojiri2007, Huang2014}. Interestingly, extended models replacing $R$ by a non-linear function $f(R)$ in the action \cite{Sotiriou2008, DeFelice:2010aj, Faraoni:2010pgm} (see also Refs.~\cite{Capozziello:2011et, Nojiri2011, Clifton:2011jh, Nojiri:2017ncd}) produce results that GR alone (i.e, without a Cosmological Constant and an unknown inflaton field) is unable to obtain \cite{Capozziello:2019cav}. 

The simplest choice, $f(R)= R+ \alpha R^2$, known as the Starobinsky model \cite{Starobinsky1980}, was shown to be in agreement with the {\it Planck} 2018 data for the inflationary epoch via the analysis of cosmic-microwave-background anisotropies \cite{Planck:2018jri}. However, the $f(R)$ model should also confront astrophysical measurements, such as the structure of compact stars, especially in cases where GR has difficulties in providing a good description \cite{Olmo2020}. For instance, the compact object of mass $2.59^{+0.08}_{-0.09}~M_{\odot}$ detected in the GW190814 event by LIGO/Virgo \citep{LIGOScientific:2020zkf} still represents a theoretical challenge to GR (assuming it to be composed of hadronic matter only \cite{Tsokaros:2020hli}). From the standpoint of $f(R)$ gravity, such object is not a surprise but a prediction \cite{Astashenok2020}. Similar results have also been found assuming non-minimal coupling between gravity and matter but otherwise keeping the linear Hilbert-Einstein Lagrangian \cite{Bertulani}.

In the last decade, compact stars were studied in $f(R)$ theories of gravity in the metric \cite{Cooney2010, Arapoglu:2010rz, Orellana2013, Astashenok2013, Yazadjiev2014, Ganguly2014, Astashenok:2014dja, Yazadjiev2015, Resco2016, Astashenok2017, Sbisa2020, Astashenok2020, Pretel2020JCAP, Panotopoulos2021} and Palatini \cite{Wojnar2018, Wojnar2019, Silveira2021, Herzog:2021wpj} frameworks. Within the metric formalism, the stellar structure equations are solved either perturbatively, i.e.~$f(R)$ is considered a small perturbation from GR \cite{Cooney2010, Arapoglu:2010rz, Orellana2013, Astashenok2013, Resco2016}, or non-perturbatively, where the full fourth-order differential equations have to be solved self-consistently for the interior and exterior of the star with suitable boundary conditions \cite{Yazadjiev2014, Ganguly2014, Astashenok:2014dja, Yazadjiev2015, Sbisa2020, Astashenok2020, Pretel2020JCAP}. The main consequence of including the $\alpha R^2$ term is, in the non-perturbative case, an increase in the value of the maximum mass (i.e., the asymptotic mass measured at infinity) of compact stars for positive values of $\alpha$. However, using the perturbative approach one finds that the maximum mass decreases as $\alpha$ increases (see e.g. Ref.~\cite{Orellana2013}). The reason for such discrepancy is that in the non-perturbative method an extra mass contribution appears in the outer region of the star due to $R \neq 0$.

Taking advantage of the fact that $f(R)$ theories are mathematically equivalent to the Brans-Dicke theory, in Ref.~\cite{Staykov2014} the authors investigated self-consistently slowly rotating neutron and strange stars in $f(R)= R+ \alpha R^2$ gravity, where they found that the neutron star moment of inertia for large values of the parameter $\alpha$ can be up to $30\%$ larger compared to the corresponding pure GR models. Through the equations governing the non-radial oscillations of spherically symmetric compact stars in the Cowling approximation within the framework of $R$-squared gravity, it was showed that the observed maximum deviation between the $f$-mode frequencies in Einstein gravity and $R^2$-gravity is up to $10\%$ and depends on the value of the free parameter $\alpha$ \cite{Staykov2015}. Furthermore, in the same modified gravity theory, the orbital and the epicyclic frequencies of a particle moving on a circular orbit around neutron or strange stars were examined in Ref.~\cite{Staykov2015EPJC}.

In this work we explore the stellar structure of interacting-quark stars using the Starobinsky model to derive the analog of the Tolman-Oppenheimer-Volkov (TOV) equations for hydrostatic equilibrium but with boundary conditions adequate to this particular modified-gravity model. Such exotic stars were previously studied using the simplest version of the MIT bag model \cite{Astashenok:2014dja,Astashenok2017}, where strong interactions are essentially ignored.  However, calculations within cold and dense perturbative quantum chromodynamics (pQCD) \citep{Kapusta2006} have shown that the short-range strong interactions have sizeable effects at intermediate and high densities, which are the conditions of interest for the interior of quark stars and neutron stars \cite{Blaizot:2000fc,Fraga:2001id,Fraga:2004gz,Kurkela:2009gj,Fraga:2013qra,Kurkela:2014vha,Fraga:2015xha,Ghisoiu:2016swa,Annala:2017llu,Gorda:2018gpy,Annala:2019puf,Gorda:2021kme}. For our calculations we use the three-loop result of Ref.~\cite{Kurkela:2009gj}, but in a more amenable analytic form \cite{Fraga:2013qra}. The results obtained are parametrized only by the pQCD dimensionless renormalization scale $X$. 

This work is organized as follows. In Sec. \ref{Sec2} we summarize the main ideas behind the perturbative QCD equation of state for cold quark matter. Section \ref{Sec3} explains some technical details related to the structure equations for compact stars within the Starobinsky model. In Sec. \ref{Sec4} we present our numerical results for the masses, radii and total binding energies of quark stars. Finally, Sec. \ref{Conclusion} summarizes our main findings and future perspectives.


\section{Cold Quark Matter from pQCD}\label{Sec2}
The equation of state (EoS) for a system composed by up, down, and strange quarks at zero temperature, which we will refer to as cold quark matter, was obtained within pQCD more than four decades ago by Freedman and McLerran \cite{Freedman:1976ub,Freedman:1977gz}, and also by Baluni \cite{Baluni:1977ms} and Toimela \cite{Toimela:1984xy}. Since then, it has been systematically improved (see, e.g. \cite{Blaizot:2000fc,Fraga:2001id,Fraga:2004gz,Kurkela:2009gj,Fraga:2013qra,Kurkela:2014vha,Fraga:2015xha,Ghisoiu:2016swa,Annala:2017llu,Gorda:2018gpy,Annala:2019puf,Gorda:2021kme}). 

In this paper we use the three-loop result of Ref.~\cite{Kurkela:2009gj}, which also includes the running of the strong coupling and strange quark mass using renormalization-group equations. For numerical convenience, we take advantage of the fact that their result can be cast in the following simple pocket formula for the pressure \cite{Fraga:2013qra}
\begin{equation}
p = p_{\rm SB}(\mu_{B})\left(c_{1}-\frac{a(X)}{(\mu_{B}/{\rm GeV})-b(X)}\right)  \; ,
\label{pressureFKV}
\end{equation}	
where $p_{\rm SB}(\mu_{B})=(3/4\pi^{2})(\mu_{B}/3)^{4}$ is the pressure of a Stefan-Boltzmann gas of $N_{f}=3$ massless quarks, $\mu_{B}$ is the baryon chemical potential, and $X=3\bar{\Lambda}/\mu_{B}$ is the dimensionless renormalization scale parameter\footnote{The  renormalization scale $\bar{\Lambda}$ naturally appears in dimensional-regularization calculations in the $\overline{\rm MS}$ scheme. The eventual dependence of physical observables on this scale is an artifact of the truncation of the perturbative series \cite{Kapusta2006}.}. The fitting functions in this formula are defined as
\begin{equation}
	a(X)=d_{1}X^{-\nu_{1}},\hspace{0.5cm}b(X)=d_{2}X^{-\nu_{2}},
\end{equation}		
with (see Ref.~\cite{Fraga:2013qra} for further details)
\begin{equation}
	c_{1}=0.9008,\hspace{0.2cm}d_{1}=0.5034,\hspace{0.2cm}d_{2}=1.452,
\end{equation}
\begin{equation}
	\nu_{1}=0.3553, \hspace{0.3cm}\nu_{2}=0.9101.
	\end{equation}
For brevity, we refer to this pocket formula as FKV[$X$].  It adds the contributions of massless up and down quarks, a strange quark with a non-zero running mass, and massless electrons that are included to ensure $\beta$-equilibrium and electric charge neutrality. 

\begin{figure}
\begin{center}
 \includegraphics[width=8.6cm]{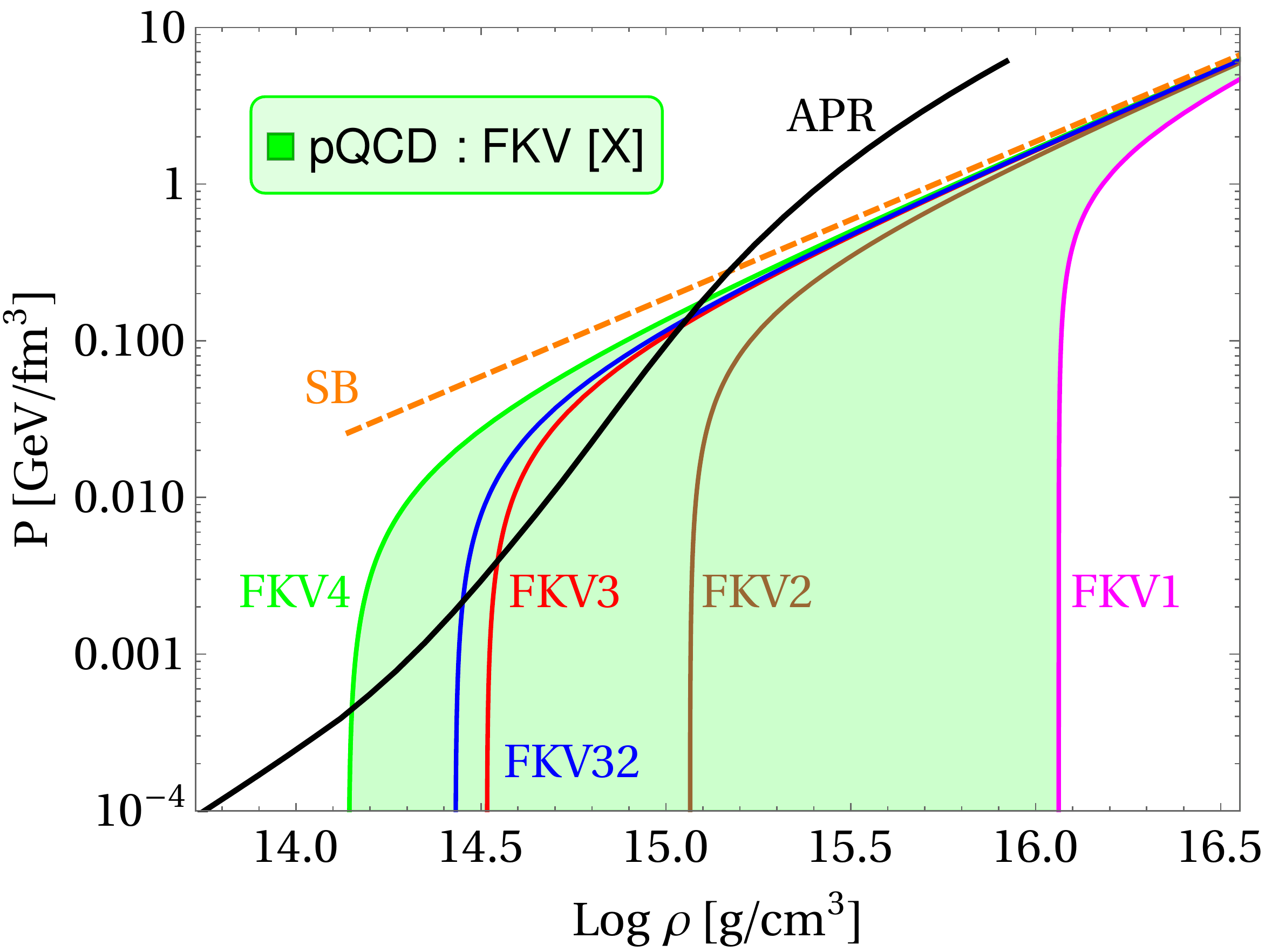}
 \caption{\label{figureEoSs} Equations of state for cold quark matter from pQCD, the Stefan-Bolztmann limit (${\rm SB}$) and hadronic matter (APR \cite{Akmal:1998cf}). Notice the partial self-bound behavior of the FKV$[X]$ pressure.}
 \end{center} 
\end{figure}

Since the dimensionless renormalization scale $X$ is usually taken to vary between $1$ and $4$ (see e.g. Ref. \cite{Kurkela:2009gj} for a discussion), Eq. (\ref{pressureFKV}) provides a band of EoSs parametrizing our ignorance of the non-perturbative QCD physics occurring at low densities and serves as a measure of the uncertainty in the calculation. This band shrinks at high densities, as required by asymptotic freedom (see Fig. \ref{figureEoSs}). 

A comment about the selfboundness of the FKV pressure is in order. By noticing that Eq. (\ref{pressureFKV}) can take the form of a bag-like EoS using standard thermodynamics, i.e.
\begin{equation}
    p = \frac{1}{3}\left(\rho - t^{\mu}_{\mu} (\mu_{B}, X) \right) \; ,
    \label{baglike}
\end{equation}
where $\rho$ is the energy density and $t^{\mu}_{\mu}$ is the trace anomaly normalized by $p_{\rm SB}$
   \begin{equation}
   t^{\mu}_{\mu}(\mu_{B}, X)=\frac{\mu_{B}}{\rm GeV}\frac{a(X)}{\left[(\mu_{B}/{\rm GeV})-b(X)\right]^{2}} \; ,
   \label{trace}
   \end{equation}
which measures the breaking of conformal symmetry brought about by the short-range QCD interactions; one can easily deduce from Eq. (\ref{baglike}) the selfbound behavior of this EoS for cold quark matter from pQCD, i.e. zero pressure at a finite energy density of $\rho_{0}({\rm pQCD})=t^{\mu}_{\mu}(\mu^{0}_{B}[X],X)$ where $\mu^{0}_{B}[X]$ is the baryochemical potential at $p=0$ for a given $X$. Notice that in the usual MIT bag model $\rho_{0}({\rm MIT})=4B$ \cite{Alcock:1986hz}. Besides, a rapid inspection of Fig. \ref{figureEoSs} indicates that larger $\rho_{0} ({\rm pQCD})$ are obtained for low values of $X$, e.g. between 1 and 2. This automatically ensures the selfboundness property. On the other hand, large $X$ tend to produce $\rho_{0} ({\rm pQCD})\to 0$, i.e. the selfboundness is lost and the EoS behaves more like a neutron-star matter one for which zero pressure is reached at a vanishing energy density.

In what follows we explore the available parameter space for $X\in [1,4]$, which allows for two classes of stars. First, there are strange stars satisfying the Bodmer-Witten hypothesis of $\rho/n_{B} \leq 0.93~{\rm GeV}$ in the vacuum ($p=0$). That happens for $2.92 \leq X \leq 4 $. Second, quark stars. Those would, in principle, have a hadronic mantle that we disregard for simplicity, leaving its inclusion for a future study. We stress that although the strange quark matter hypothesis is realized for quark stars having $X \gtrsim 3$, our above analysis shows that they are not strictly selfbound and gravity is the dominating binding interaction. We will see that this property is manifested through the behavior of the associated total binding energies.


\section{Modified gravity TOV equations}\label{Sec3}
The action for $f(R)$ modified theories of gravity in the Jordan frame is given by
\begin{equation}\label{action}
    I = \frac{1}{16\pi}\int d^4x\sqrt{-g}f(R) + I_m ,
\end{equation}
where $g$ is the determinant of the metric tensor $g_{\mu\nu}$, $R$ the Ricci scalar curvature and $I_m$ stands for the matter action. 
We choose the metric formalism (as opposed to the Palatini approach, which assumes that the metric $g_{\mu\nu}$ and the connection $\Gamma^\alpha_{\mu\nu}$ are independent objects) \cite{Koivisto2006}, obtaining \cite{Sotiriou2008, DeFelice:2010aj, Faraoni:2010pgm}
\begin{equation}\label{FieldEq}
f_R R_{\mu\nu} - \dfrac{1}{2}g_{\mu\nu}f + [g_{\mu\nu}\square - \nabla_\mu\nabla_\nu] f_R  = 8\pi T_{\mu\nu} ,
\end{equation}
with $T_{\mu\nu}$ being the matter energy-momentum tensor, $f_R(R) \equiv df(R)/dR$, $\nabla_\mu$ is the covariant derivative associated with the Levi-Civita connection, and $\square \equiv \nabla_\mu \nabla^\mu$ is the d'Alembertian operator. 

Besides, notice that in any  $f(R)$ model the Ricci scalar is governed by a second-order differential equation obtained from the trace of the field equation:
\begin{equation}\label{RicciS}
    3\square f_R(R) + Rf_R(R) - 2f(R) = 8\pi T ,
\end{equation}
where $T \equiv g_{\mu\nu}T^{\mu\nu}$. As expected, Eq. (\ref{RicciS}) reduces to $R = -8\pi T$ in the GR limit where it is just an algebraic equation. Nonetheless, for nonlinear functions in $R$, both the metric and the Ricci scalar are dynamical fields, i.e. described by differential equations \cite{Olmo2007}. As a consequence, in $f(R)$ theories of gravity, $T= 0$ no longer implies that $R=0$ or, in other words, outside the matter-energy source there are still persistent curvature effects.

For our study of compact stars, we consider a static and spherically symmetric system, so that the spacetime element can be written in the standard form
\begin{equation}\label{LineEle}
    ds^2 = -e^{2\psi(r)}dt^2 + e^{2\lambda(r)}dr^2 + r^2(d\theta^2 + \sin^2\theta d\phi^2) ,
\end{equation}
and the matter-energy distribution is described by an isotropic perfect fluid with energy density $\rho$, pressure $p$ and four-velocity $u^\mu$, i.e.
\begin{equation}\label{EnerMomT}
    T_{\mu\nu} = ( \rho + p )u_\mu u_\nu + pg_{\mu\nu} . 
\end{equation}

We can derive the modified TOV equations within the $f(R)= R+ \alpha R^2$ gravity model\footnote{Here $\alpha$ is a free parameter having units of $r_g^2$, where $r_g = GM_\odot/c^2 \approx 1.477\ \text{km}$ is the solar mass in geometrical units.} using Eqs. (\ref{FieldEq})--(\ref{EnerMomT}) along with the generalized Bianchi identity \cite{Koivisto2006}, which expresses the energy-momentum conservation, obtaining \citep{Yazadjiev2015, Pretel2020JCAP}
%
\begin{align}
    \frac{d\psi}{dr} &= \dfrac{1}{4r( 1+ 2\alpha R + \alpha rR' )}\left[ r^2e^{2\lambda}(16\pi p - \alpha R^2) + 2(1 + 2\alpha R)\left( e^{2\lambda} -1 \right) - 8\alpha r R' \right] , \label{TOV1}  \\
    \frac{d\lambda}{dr} &= \dfrac{1}{4r(1+ 2\alpha R + \alpha rR')}\left\lbrace 2(1 + 2\alpha R)\left(1-e^{2\lambda}\right)  + \dfrac{r^2e^{2\lambda}}{3}\left[ 16\pi (2\rho + 3p) + 2R + 3\alpha R^2  \right]\right.   \nonumber  \\
    & \hspace{0.08\textwidth}\left. + \dfrac{2\alpha rR'}{1+2\alpha R}\left[ 2(1 + 2\alpha R)\left(1-e^{2\lambda}\right) + \dfrac{r^2e^{2\lambda}}{3}(16\pi\rho + R + 3\alpha R^2) + 4\alpha rR' \right]\right\rbrace , \label{TOV2}   \\  
    \frac{d^2R}{dr^2} &= \dfrac{e^{2\lambda}}{6\alpha}\left[ R + 8\pi (3p - \rho) \right] + \left( \lambda' - \psi' - \dfrac{2}{r} \right)R' ,  \label{TOV3}  \\
    \frac{dp}{dr} &= -(\rho + p)\psi' ,   \label{TOV4}
\end{align}
%
where the prime denotes derivative with respect to the radial coordinate $r$. As in GR, the stellar surface is found when the pressure vanishes, i.e. $p(r= r_{\rm sur})= 0$.

In Eqs. (\ref{TOV1})-(\ref{TOV4}) we have a system of three first-order and one second-order differential equations with a set of five functions to be determined: $\psi$, $\lambda$, $R$, $\rho$ and $p$ . As in the case of Einstein's gravity, the EoS for dense matter, $p= p(\rho)$, must be included so that only five boundary conditions are required to solve this system inside the star. Indeed, by ensuring regularity of the geometry at the stellar center, we can establish the following boundary conditions
\begin{align}\label{BC1}
    \rho(0) &= \rho_c ,   &   \psi(0) &= \psi_c ,   &    \lambda(0) &= 0 ,  \nonumber   \\
    R(0) &= R_c ,   &   R'(0) &= 0 ,
\end{align}
where $\rho_c$ and $R_c$ are the values of the central energy density and central scalar curvature, respectively. Besides, the solution outside the star is defined by Eqs. (\ref{TOV1})--(\ref{TOV3}), where both density and pressure vanish ($\rho =p =0$). At the stellar surface, it is therefore useful to settle the junction conditions as  
\begin{align}\label{BC2}
    \psi_{in} (r_{\rm sur}) &= \psi_{out} (r_{\rm sur}) ,   &   \lambda_{in} (r_{\rm sur}) &= \lambda_{out} (r_{\rm sur}) ,  \nonumber  \\
    R_{in} (r_{\rm sur}) &= R_{out} (r_{\rm sur}) ,   &   R'_{in} (r_{\rm sur}) &= R'_{out} (r_{\rm sur}) .
\end{align}

Here it is convenient to define a mass function for the stellar system. To do that, we use the $00$-component of the field equations, which can be recast in the form
\begin{align}\label{ttComponentFE}
    \frac{d}{dr}\left( re^{-2\lambda} \right) =&\ 1- 8\pi r^2\rho - 2\alpha \left\lbrace - R\frac{d}{dr}\left[ r(1- e^{-2\lambda}) \right] \right.  \nonumber  \\
    &\left. + \frac{r^2}{4} R^2  + \frac{r^2}{e^{2\lambda}}\left[ \left( \frac{2}{r}- \lambda' \right)R'+ R'' \right]  \right\rbrace ,
\end{align}
i.e., the metric function $\lambda$ is generated by the matter fields and the terms related to the Ricci scalar. The asymptotic (weak-field) limit implies that one can define a so-called ``astrophysical mass parameter" $m(r)$ by
\begin{equation}\label{ExpLambda}
    e^{-2\lambda} = 1 - \frac{2m}{r} .
\end{equation}
There are, however, different (unequivalent) definitions of ``mass" in $f(R)$ theories \cite{Sbisa2020}. This means that we can define a mass parameter in $R$-squared gravity, just as in Einstein's theory, which characterizes the mass enclosed within the radius $r$, given by
\begin{align}
    m =&\ 4\pi \int r^2\rho dr + \alpha\int\left\lbrace \frac{R^2}{4} - \frac{R}{r^2}\frac{d}{dr}\left[ r(1- e^{-2\lambda}) \right]  \right.  \nonumber  \\
    &\left. + \frac{1}{e^{2\lambda}}\left[ \left( \frac{2}{r}- \lambda' \right)R' + R'' \right] \right\rbrace r^2dr .  \label{MassFuncEq}
\end{align}

In fact, when $\alpha = 0$, the second rhs term vanishes and we retrieve the well-known expression in GR where there is no contribution to the mass from regions outside the star itself. Here, with $\alpha\neq{0}$, the scenario is different: even in the outer region of a compact star (where $\rho = p =0$), Eq. (\ref{MassFuncEq}) generates an extra mass contribution due to the Ricci scalar and its derivatives\footnote{Some authors have introduced the term ``gravitational sphere'' to refer to the amount of mass that surrounds the star \cite{Astashenok:2014dja,Astashenok2017}.}. 

Notice that Eq. (\ref{ExpLambda}) puts additional constraints coming from the  asymptotic flatness requirement on the Ricci scalar and the mass parameter, i.e.:
\begin{align}\label{AsyFR}
    \lim_{r \rightarrow \infty} R(r) &= 0 ,   &   \lim_{r \rightarrow \infty} m(r) = \rm constant .
\end{align}
In other words, $R_c$ must be chosen appropriately to satisfy Eqs. (\ref{AsyFR}) at infinity. At the same time, the central value of the metric function $\psi$ in Eq. (\ref{BC1}) is fixed by requiring asymptotic flatness, namely $\psi(r\rightarrow \infty) \rightarrow 0$. Thus, the total gravitational mass of the star $M$ is determined from the asymptotic behavior (see Eq. (\ref{ExpLambda}))
\begin{equation}\label{MasstotEq}
    M \equiv \lim_{r \rightarrow \infty} \frac{r}{2}\left( 1- \frac{1}{e^{2\lambda}} \right) .
\end{equation}

Finally, a comment about the values taken by $\alpha$ is in order. Current constraints from mass-radius diagrams (see, for instance, Ref.~\cite{Arapoglu:2010rz} for neutron stars) put an upper limit at about $\alpha_{\rm max} \sim 0.5 r^{2}_{g}$ \cite{Naf:2010zy}. Nevertheless, the reason our work considers also larger values, i.e. $\alpha<10r^{2}_{g}$, is threefold: First, the constraints could be different (i.e, weaker) from quark stars, since they follow a qualitatively different EoS. Secondly, we still wish to compare our findings with previous results in the literature (see, e.g. \cite{Astashenok2013,Astashenok2017}). Finally, the adopted range is the narrower one that still allows a visual gap between the different curves.

%
\begin{figure}
 \includegraphics[width=7.7cm]{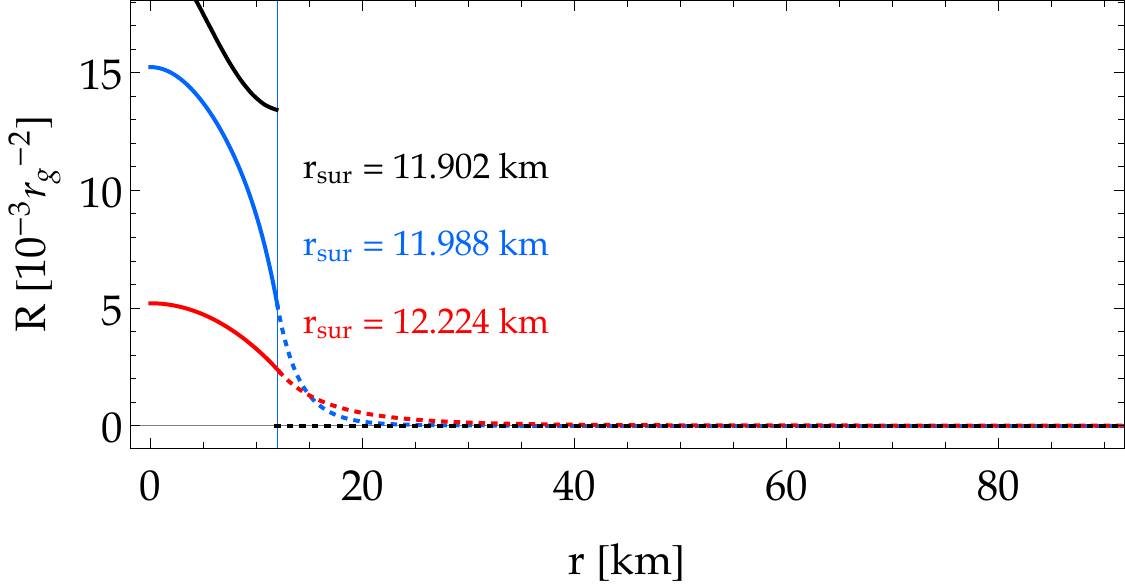}\
 \includegraphics[width=7.7cm]{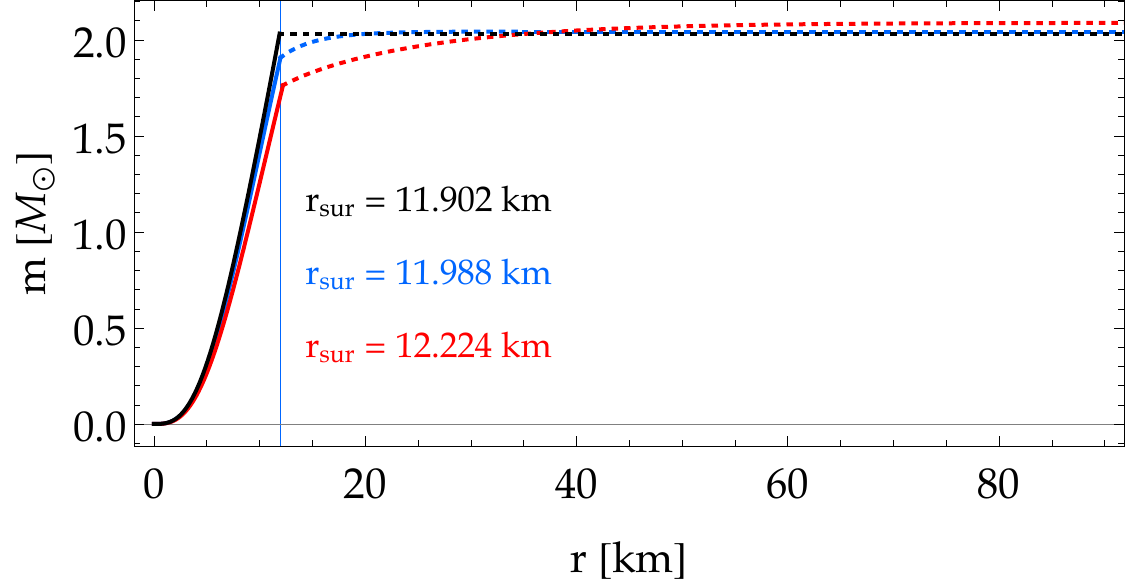}
 \caption{\label{figureRm} Ricci scalar (left panel) and mass function (right panel) profiles for a quark star obtained from the FKV3 EoS with central density $\rho_c = 1.5 \times 10^{15} \rm g/cm^3$ within the Starobinsky model with $\alpha = 1 r_g^2$ (blue curves) and $\alpha = 10 r_g^2$ (red curves). For $r>r_{\rm sur}$, both Ricci scalars go smoothly to zero at infinity in agreement with the asymptotic flatness requirement, the mass function approaching a constant value given by $M = 2.044 M_\odot$ (for $\alpha = 1 r_g^2$) and $M = 2.089 M_\odot$ (for $\alpha = 10 r_g^2$). The solid and dotted curves correspond to the interior and exterior solutions, respectively. The GR case is shown by black lines, where the total mass is $M = 2.033 M_\odot$.}
\end{figure}
%

\section{Numerical Results} \label{Sec4}

We can now solve numerically the modified-gravity TOV equations (\ref{TOV1})--(\ref{TOV4}), with boundary conditions given by Eqs. (\ref{BC1})--(\ref{BC2}). The EoS given by FKV$[X]$ generates quark stars or strange stars depending on the value of $X$, as discussed in Sec. \ref{Sec2}. For comparison, we also show results for the APR EoS \cite{Akmal:1998cf}, since it is widely used in dense nuclear matter calculations (see Fig. \ref{figureEoSs}). Each equation of state will provide an input to the modified-gravity equations of hydrostatic equilibrium.

%
\begin{figure}
 \includegraphics[width=7.7cm]{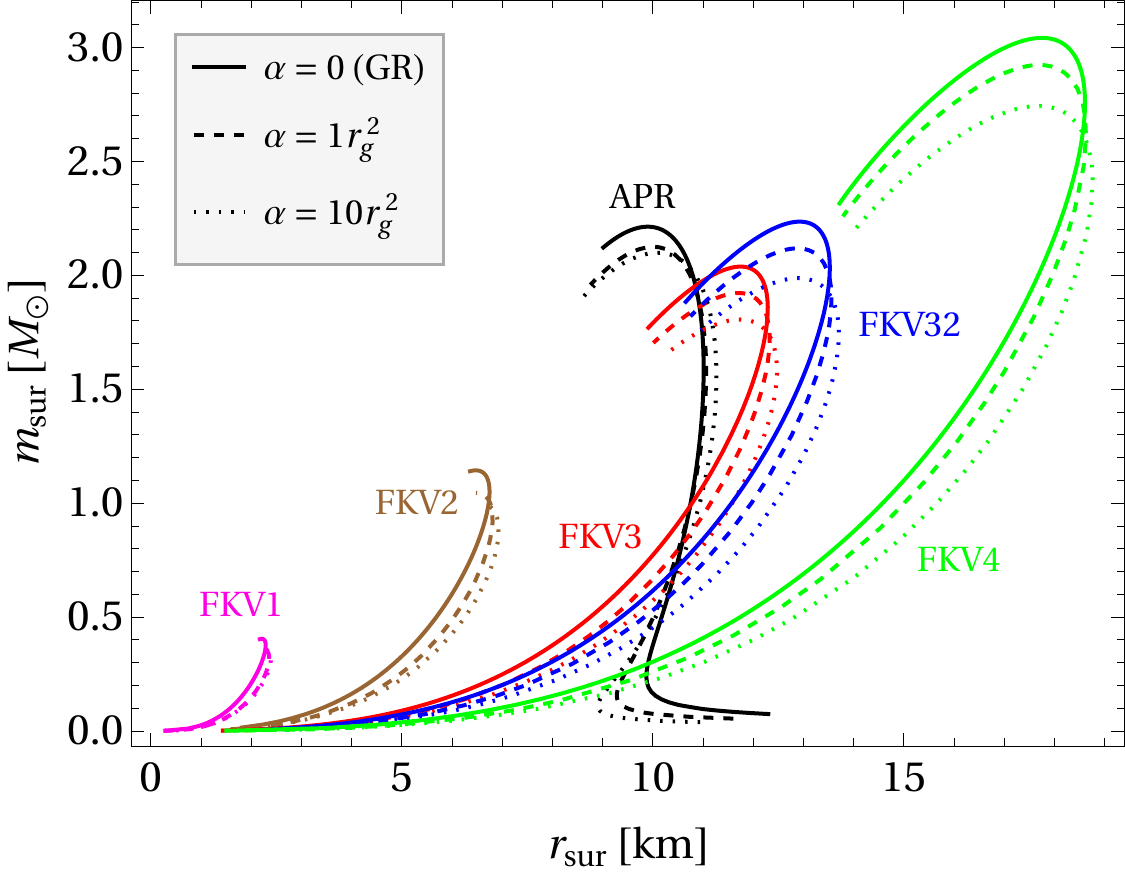}\
 \includegraphics[width=7.7cm]{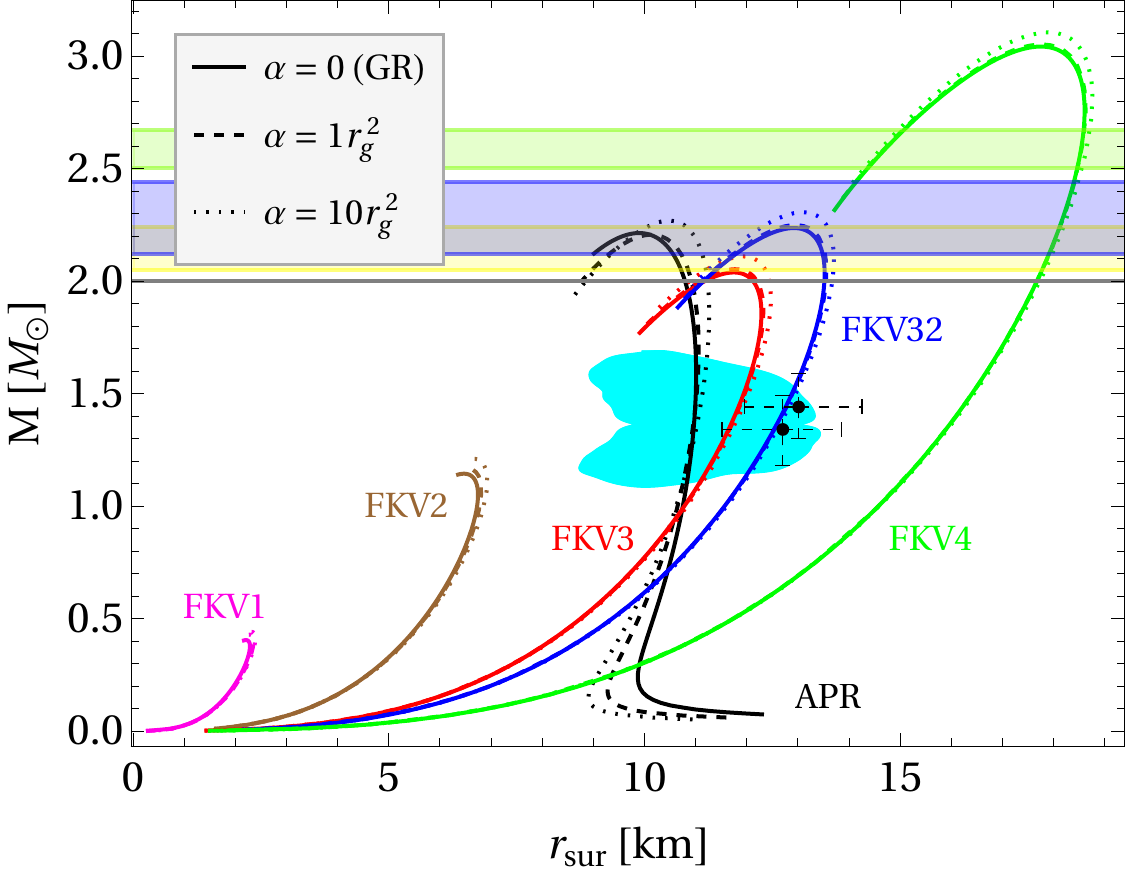}
 \caption{\label{figureMassRadius} Gravitational mass at the stellar surface, $m_{\rm sur}$, versus the surface radius $r_{\rm sur}$ (left panel) and total mass at infinity, $M$, versus the surface radius (right panel) in the Starobinsky gravity model for three moderate values of $\alpha$ and different equations of state. The gray line at $2.0 M_\odot$ represents the two massive NS pulsars J1614-2230 \cite{Demorest:2010bx} and J0348+0432 \cite{Antoniadis:2013pzd}. The yellow and blue horizontal bands (partially overlapping each other) stand for the observational measurements of the masses of the highly massive NS pulsars J0740+6620 \cite{NANOGrav:2019jur} and J2215+5135 \cite{Linares:2018ppq}, respectively. The filled green stripe represents the lower mass of the compact object detected by the GW190814 event \citep{LIGOScientific:2020zkf}, and the filled cyan region is the mass-radius constraint from the GW170817 event. The NICER measurements for PSR J0030+0451 are indicated by black dots with their respective error bars \cite{Miller:2019cac, Riley:2019yda}. }  
\end{figure}
%

The stellar structure of the quark star families is obtained, as usual, by mapping out different values of the central mass density into the EoS, i.e. $p(r=0)\equiv{p}_{c}\equiv{p}(\rho_{c})$. But now we must also choose central values of the Ricci scalar $R_c$ that fulfills  the asymptotic-flatness requirement, $R(r\rightarrow\infty) \rightarrow 0$, for the exterior solution, i.e. $R_{c}$ exhibits a unique value for each stellar configuration. For example, we show in Fig. \ref{figureRm} the interior and exterior behavior of the Ricci scalar and mass function for a strange star obtained from the FKV$[3]$ EoS with central mass density $\rho_c = 1.5 \times 10^{15}\ \rm{g}/\rm{cm}^3$ for two values of the Starobinsky parameter $\alpha$, as well as the standard GR results (which, outside the star, obviously vanishes in the left panel and is exactly constant in the right one). As explained in Sec. \ref{Sec3}, finite values of $R$ in the star exterior generate an extra mass contribution according to the second integral in the rhs of Eq. (\ref{MassFuncEq}). The mass function takes a constant value (up to $0.01\%$) at spatial infinity, thus becoming the usual gravitational mass $M$, as expected.

In Fig. \ref{figureMassRadius} we display our findings for the mass parameter at the star surface, $m_{\rm sur}\equiv~m(r_{\rm sur})$, versus the surface radius, $r_{\rm sur}$, where only the pressure goes to zero. One can see that $m_{\rm sur}$ decreases significantly compared to GR (dropping by $\sim 0.4~ M_{\odot}$ in some cases) for moderate values of $\alpha$. This can be understood by noticing that the Ricci scalar contributes less than its total value up to $r_{\rm sur}$, e.g. $\sim{3R}/4$ for $\alpha=10r^{2}_{g}$ (see Fig. \ref{figureRm}), the remaining part of $R$ contributing only to the asymptotic mass, $M\equiv~m(r\to \infty)$, thus making it higher. Furthermore, notice that these increments are negligible for low- and intermediate-mass stars and only noticeable for the heaviest ones associated to each FKV$[X]$ EoS where the deviations from GR reach\footnote{Another interpretation comes from Eq. (\ref{RicciS}). It shows that, in order to get modified-gravity curvature, a higher $\rho_{c}$ is needed, thereby producing more massive stars.} $0.1~M_{\odot}$ for $X>3$. Hadronic stars obtained from the APR EoS exhibit a similar qualitative behavior, but their variations in $m_{\rm sur}$ and $M$ are less than $0.1~M_{\odot}$.

%
\begin{figure*}
 \includegraphics[width=7.7cm]{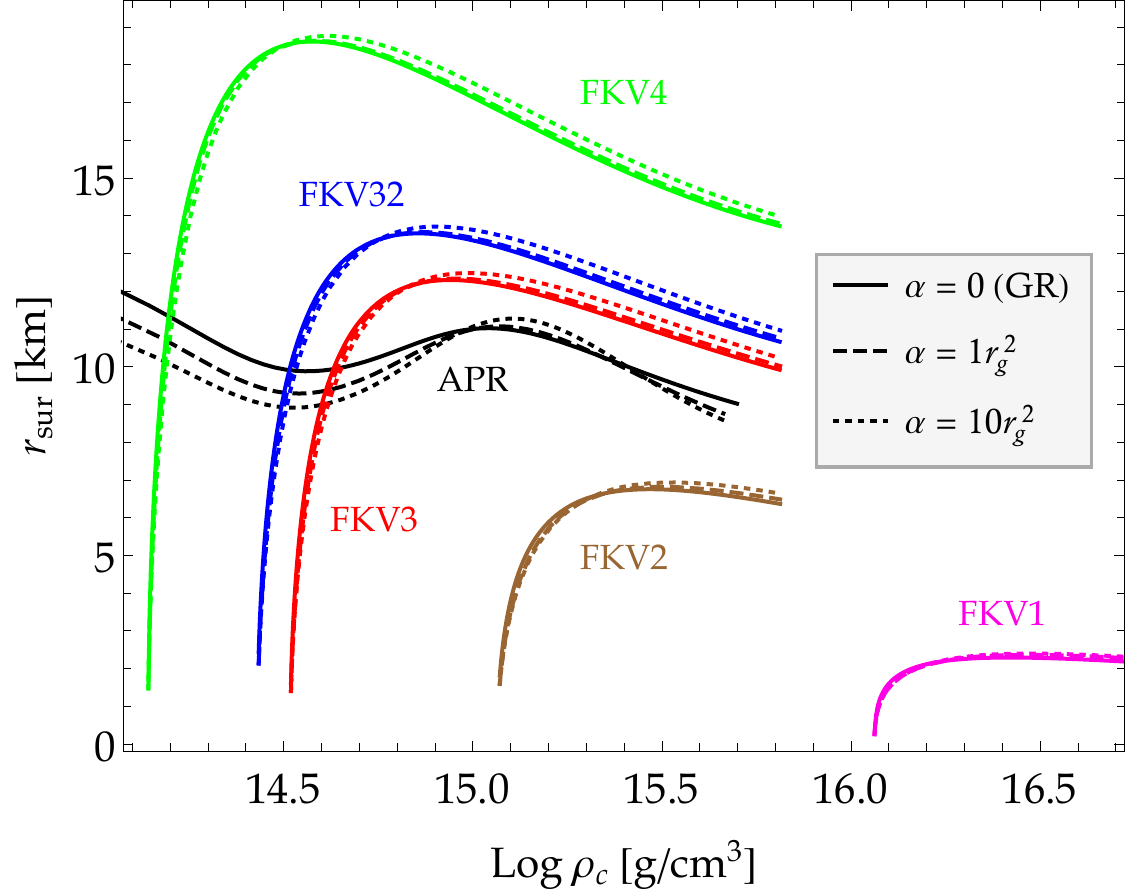}
 \includegraphics[width=7.75cm]{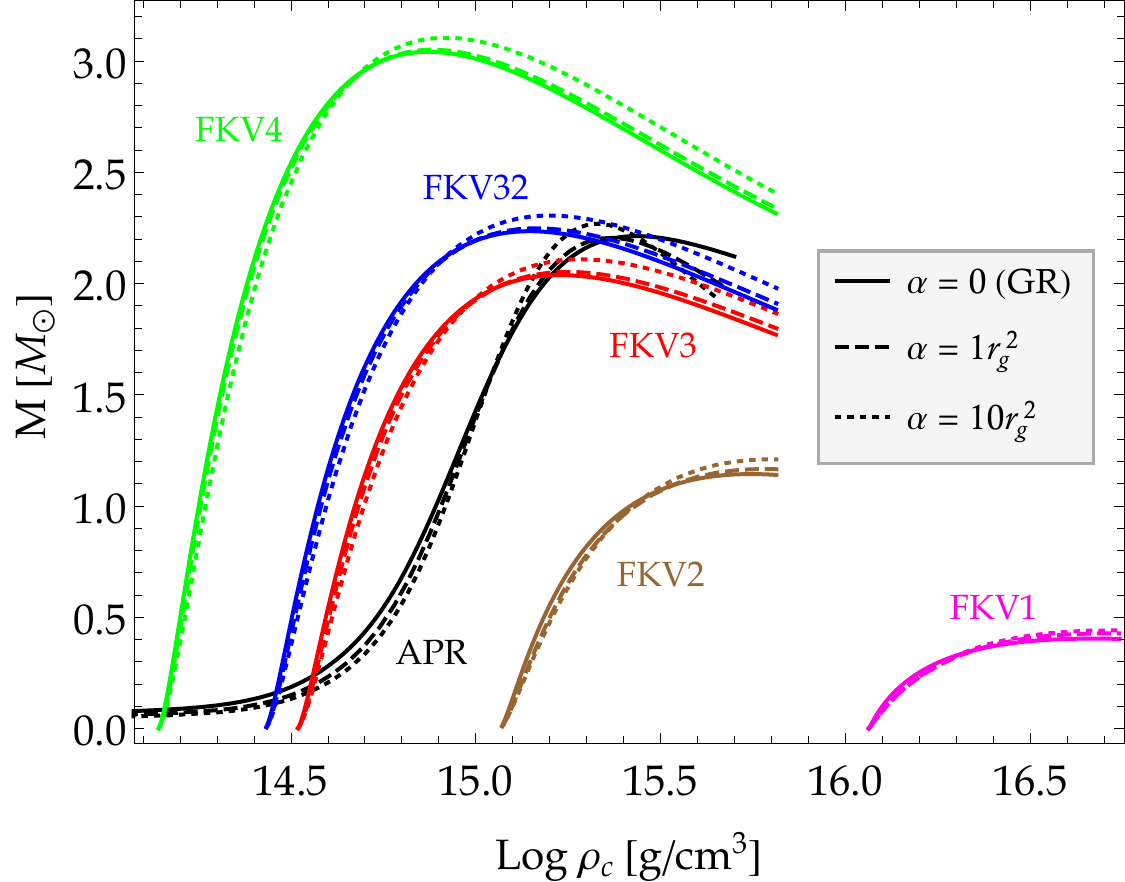}\
 \caption{\label{figureMassDen} Surface radius $r_{\rm sur}$ (left panel) and gravitational mass $M$ (right panel) vs the central mass density for quark stars and hadronic stars in GR ($\alpha=0$) and $R^{2}$-gravity ($\alpha=1~{\rm and}~10$ in units of $r^{2}_{g}$).}  
\end{figure*}
%

The dependence of the total mass, $M$, and surface radius, $r_{\rm sur}$, on the central mass densities, $\rho_{c}$, for quark stars is illustrated in Fig. \ref{figureMassDen}. One can observe the usual increasing of the total mass $M$ up to a critical density $\rho_{c}$, corresponding to the maximum mass $M_{\rm max}$. In GR, one usually considers stable quarks stars as long as they satisfy the static stability criterion $\partial{M}/\partial{\rho_{c}}\geq{0}$ \cite{Glendenning:2000}. Unfortunately, a similar condition has not been proved rigorously for the case of Starobinsky quark stars yet. Nevertheless, the GR criterion still makes sense physically, since stars with decreasing masses and large $\rho_{c}$ are doomed to undergo gravitational collapse \cite{Glendenning:2000}. 

Furthermore, notice that the effect of the $\alpha R^2$ term begins to be more appreciable only after passing the maximum mass, especially for $3 \leq X \leq 4$. In Fig. \ref{figureMassDen} one can also see that the surface radius $r_{\rm sur}$ is almost equivalent to the one in GR up to $\rho^{\rm max}_{c}$ and only slightly different afterwards, where there are only unstable stars. Besides, the mass at infinity $M$ of the hadronic stars is reduced for low-mass stars and slightly increased in the neighbourhood of the maximum mass. The impact of the $\alpha{R}^{2}$ term on the hadronic star is larger in $r_{\rm sur}$,  although mostly pronounced at low central-mass densities. 


In Table \ref{table1} we list the variations of the central Ricci scalar and central mass density for increasing values of the Starobinsky parameter $\alpha$. One can notice a reduction in non-trivial $R_{c}$ and increment of $\rho_{c}$, which barely affects $r_{\rm sur}$ but appreciably modifies $m_{\rm sur}$. On the other hand, the mass at infinity ($M$) does not change significantly. The fact that $r_{\rm sur}$ and $M$ are not significantly modified leads us to infer that observables such as the gravitational redshift $z_{\rm sur}$, given by
\begin{equation}
z_{\rm sur} +1 = \left(1- \frac{2m_{\rm sur}}{r_{\rm sur}}\right)^{-1/2},
\end{equation}
will not be appreciably affected either, which does not ease the  recognition between GR and Starobinsky quark stars. Even more so, as we have pointed out above, $m_{\rm sur}$ is less than the (asymptotic) total mass $M$, i.e, it  yields a {\it smaller} $z_{\rm sur}$. Therefore, one {\it cannot} hope to obtain observational constraints on $\alpha$ via violations of the maximum surface redshift $z_{\rm max} = 2$ predicted by GR \cite{Wald}. Nevertheless, we should point out that the mass estimates from the surface redshift ($m_{\rm sur}$) and from a companion star ($M$) are different in such a theory (recall Fig. \ref{figureRm}, right panel); it remains to be determined the actual observational precision required to verify so.

Another interesting observable that can be extracted from the modified-gravity framework is the total binding energy, $E_{B}$, for a given family of quark stars. It is defined\footnote{This definition is standard in modified-gravity theories (see, e.g. Ref. \cite{Doneva:2020ped}) but differs by a minus sign in GR \cite{Glendenning:2000,Haensel:2007yy}.} as $E_B = M - M_{\rm pr}$, i.e. the difference between the asymptotic mass $M$ and the proper mass $M_{\rm pr}$ (also known as baryon mass), being the latter obtained from \cite{Doneva:2013qva}
\begin{equation}\label{ProperMass}
M_{\rm pr} = {m_{B}}N_{B} = 4\pi{m_{\rm B}}\int_0^{r_{\rm sur}} e^{\lambda(r)} r^{2}n_{B}(r)dr , 
\end{equation}
where $n_{B}(r)$ is the baryon number density profile and $m_{\rm B}$ the neutron mass being chosen to agree with different stellar evolution models requiring the conservation of baryon number, $N_{B}$, thus producing low- and high-mass quark stars through, e.g. gravitational collapse of a massive dying star \cite{Glendenning:2000,Haensel:2007yy} for the last case and the burning of hadronic stars \cite{Berezhiani:2002ks,Drago:2004vu,Bombaci:2004mt} for the former case. Notice that the total $E_{B}$ accounts for the microphysical interactions together with the gravitational contribution having positive and negative values, respectively \cite{Glendenning:2000}. If one of them dominates over the other, its sign of binding energy is kept along all the sequence of stars. By using this reasoning one can anticipate that selfbound stars (low $X$) will have $E_{B}>0$ since the short-ranged strong interactions dominate over gravity which is considered through the modified-gravity TOV equations. On the other hand, quark stars with large $X$ will have their $E_{B}<0$, i.e. it is dominated by the gravitational binding. Stellar configurations satisfying this criterium are said to be stable against baryon evaporation since they are precluded from escaping to infinity.

%
\begin{figure*}
\begin{center}
 \includegraphics[width=8.6cm]{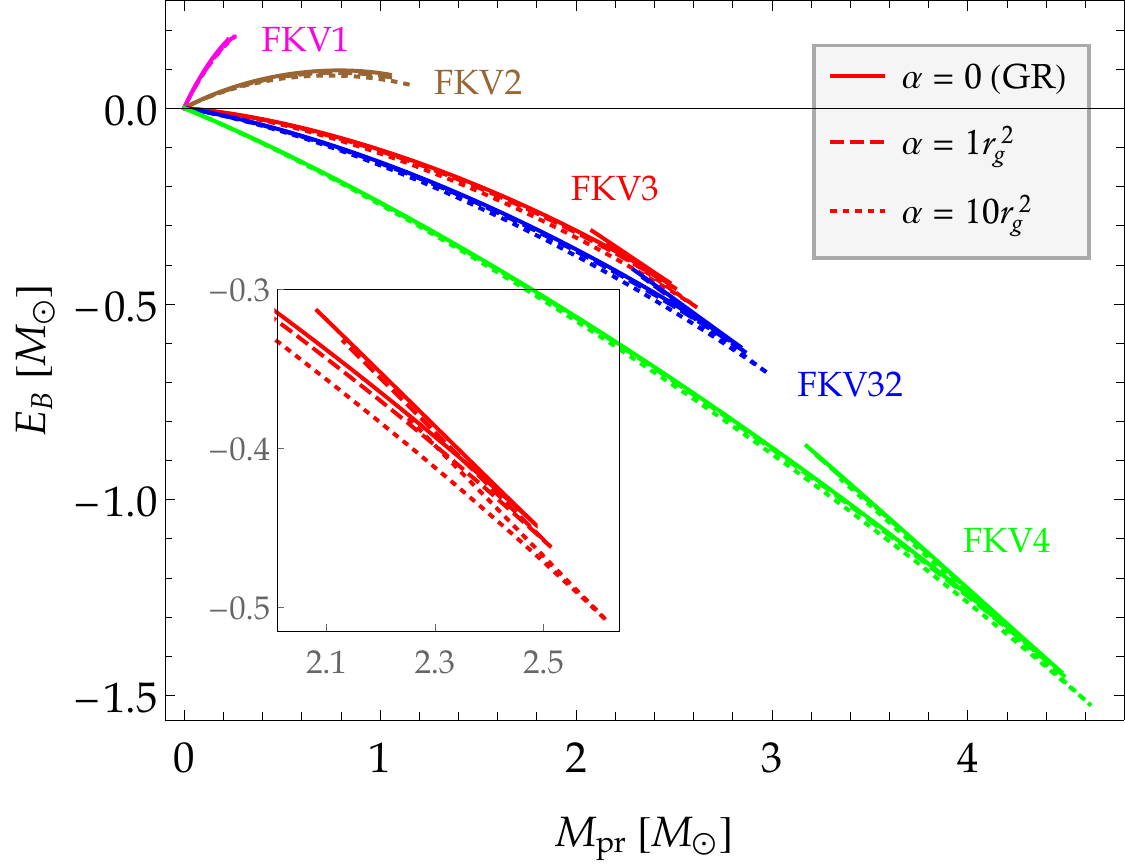}
 \caption{\label{figureBE} Gravitational binding energy $E_{B}$ as a function of the proper mass of quark stars for different values of $\alpha$. The inset magnifies the behavior for the FKV3 EoS in the surroundings of its maximum mass. Notice the rather different qualitative behavior of selfbound stars (FKV1,2) compared to the gravitationally-bound quark stars (FKV3,3.2,4). See the text for discussion.}
 \end{center}
\end{figure*}
%

%
\begin{table*}
\caption{\label{table1} 
Properties of the maximum-mass strange star configurations for three particular values of the free parameter $\alpha$. As $\alpha$ increases, the central-density value corresponding to $dM/d\rho_c=0$ is increasing, and which can be observed in the right panel of Fig. \ref{figureMassDen}.}\vspace{0.4cm}
\begin{tabular}{c|ccccccc}
EoS  &  $\alpha$ [$r^{2}_{g}$]  &  $R_c$ [$10^{-3}r^{-2}_{g}$]  &  $\rho_c$ [$10^{15} \rm g/cm^3$]  &  $r_{\rm sur}$  [\rm{km}]  &  $m_{\rm sur}$  [$M_\odot$]  &  $M$ [$M_\odot$]  &  $E_B$ [$M_\odot$]  \\
\hline
  &  0  &  21.160  &  1.693  &  11.752  &  2.037  &  2.037  &  -0.437  \\
FKV3  &  1  &  15.735  &  1.775  &  11.784  &  1.921  &  2.052  &  -0.462  \\
  &  10  &  5.284  &  1.912  &  11.939  &  1.800  &  2.109  &  -0.509  \\
\hline
  &  0  &  17.687  &  1.400  &  12.927  &  2.235  &  2.235  &  -0.612  \\
FVK32  &  1  &  13.682  &  1.459  &  12.956  &  2.117  &  2.248  &  -0.626  \\
  &  10  &  4.981  &  1.587  &  13.100  &  1.982  &  2.305  &  -0.678  \\
  \hline
  &  0  &  9.727  &  0.744  &  17.755  &  3.041  &  3.041  &  -1.443  \\
FVK4  &  1  &  8.278  &  0.763  &  17.775  &  2.922  &  3.051  &  -1.458  \\
  &  10  &  3.925  &  0.829  &  17.900  &  2.738  &  3.105  &  -1.526  \\
\end{tabular}
\end{table*}
%

In Fig. \ref{figureBE} we display our results for $E_{B}$ for different values of $X$ and the Starobinsky parameter. Notice that $E_{B}$ in the Starobinsky model behaves very similarly to the case in GR. The usual negative cusp, signaling the minimal total binding energy and associated with $M^{\rm max}$, appears for quark stars between FKV3 and FKV4. Those correspond to strange stars where both gravity and pQCD (through the Bodmer-Witten hypothesis) allow their existence in spite of not being strictly selbound (see discussion in Sec. \ref{Sec2}). The changes due to modified gravity are listed in Table \ref{table1}. On the other hand, for $X=1,2$ no cusp is formed and the binding energies are always positive, even in Einstein's gravity. This is not surprising since the corresponding quark stars are self-bound by pQCD interactions, thus not being related to any gravitational mechanism at their stellar birth \cite{Glendenning:2000}.

\section{Summary and outlook}\label{Conclusion}

In this paper we have investigated the structure of (strange) quark stars within the $R^{2}$-gravity framework, also known as the Starobinsky model, using an equation of state for quark matter obtained from cold and dense perturbative QCD. 

From the spacetime perspective, we have used a non-perturbative treatment in the metric formalism to derive the modified-gravity TOV equations which necessarily require further non-trivial boundary conditions. We have found that the mass parameter at the stellar surface, $m_{\rm sur}$, suffers more significant variations when compared to GR, whereas the deviations of the asymptotic mass, $M$, are almost negligible for small masses but more significant in the maximum-mass region. This is somewhat expected since, in this modified-gravity model, the $\alpha R^2$ term makes the Ricci scalar a dynamical field that does not vanish in the star exterior. Our calculations also show that the gravitationally-dominating ($X>2.92$) and self-boundly-dominating ($X\leq2.92$) binding energies are mainly affected by the modified-gravity term for stars with large $M_{\rm pr}$. In the case of strange stars, the cusp that marks the transition between stable/unstable stellar configurations is delayed and for self-bound stars the cusp does not exist.

Finally, we found that Starobinsky strange stars suffer minimal variations of their asymptotic maximum masses, which lie in the uncertainty error of e.g. the GW190814 compact-object mass ($\sim$~0.1 $M_{\odot}$). Thus, one is still unable to distinguish GR from modified $f(R)=R+\alpha{R}^{2}$ gravity predictions with the currently available observables of quark stars. Future perspectives include the investigation of the dynamical stability of quark stars against radial oscillations within the Starobinsky model, which are not yet fully understood.

\acknowledgments
This work was supported by INCT-FNA (Process No. 464898/2014-5). J. C. J. acknowledges support from FAPESP (Processes No. 2020/07791-7 and No. 2018/24720-6). J. M. Z. P. recognizes financial support from the PCI program of CNPq. E. S. F. is partially supported by CAPES (Finance Code 001), CNPq, and FAPERJ.


\end{document}